\documentstyle[preprint,aps,fixes,epsf]{revtex}
\begin{document}
\preprint{\hfill{\parbox{4.0cm}{SNUTP 97-014 \ \\ hep-ph/9705284}}}
\title{Azimuthal Correlation in Lepton-Hadron Scattering \\
via  Charged Weak-Current Processes}
\author{Junegone Chay\footnote{e-mail address: {\tt
chay@kupt.korea.ac.kr}} 
and Sun Myong Kim\footnote{e-mail
address: {\tt kim@kupt2.korea.ac.kr}}} 
\address{Department of Physics,  Korea University, Seoul 136-701, 
Korea}
\date{\today}
\maketitle
\begin{abstract}
We consider the azimuthal correlation of the final-state particles in 
charged weak-current processes. This correlation provides a test
of perturbative quantum chromodynamics (QCD). The azimuthal asymmetry
is large in the semi-inclusive processes in which we identify a
final-state hadron, say, a charged pion compared to that in the
inclusive processes in which we do not identify final-state particles 
and use only the calorimetric information. In semi-inclusive processes 
the azimuthal asymmetry is more conspicuous when the incident lepton
is an antineutrino or a positron than when the incident lepton is a
neutrino or an electron. We analyze all the possible charged
weak-current processes and study the quantitative aspects of each
process. We also compare this result to the $ep$ scattering with a
photon exchange.    
\end{abstract}
\narrowtext
\section{Introduction}
The QCD-improved parton model has shown a great success in describing
high-energy processes such as deep-inelastic leptoproduction. In the
parton model we can express the cross section as a convolution of
three factors: the parton-lepton hard-scattering cross section, the
distribution function describing the partons in the initial state and
the fragmentation functions describing the distribution of final-state
hadrons from the scattered parton. The hard-scattering cross section
at parton level can be calculated at any given order in perturbative
QCD. The distribution functions and fragmentation functions themselves
cannot be calculated perturbatively but the evolution of these
functions can be calculated using perturbation theory.  

The azimuthal correlations provide a clean test of perturbative
QCD since these correlations occur at higher orders in perturbative
QCD. Georgi and Politzer\cite{georgi} proposed the azimuthal angular
dependence of the hadrons in the semi-inclusive processes $\ell + p
\rightarrow \ell^{\prime} +h +X$, where $\ell$, $\ell^{\prime}$ are
leptons, $h$ is a detected hadron. Cahn\cite{cahn} included the
contribution to the azimuthal angular dependence from the intrinsic
transverse momentum of the partons bound inside the
proton. Berger\cite{berger} considered the final-state interaction
producing a pion and found that the azimuthal asymmetry due to this
final-state interaction is opposite in sign to that due to the effects
studied by Cahn. The azimuthal asymmetries discussed by Cahn and
Berger are due to nonperturbative effects. These effects were analyzed 
at low transverse momentum.\cite{mazzanti,emc}

In the kinematic regime attainable at the $ep$ collider at HERA or in
the CCFR experiments, we expect that perturbative QCD effects will
dominate nonperturbative effects. This is the motivation for
considering the azimuthal correlation of final hadrons in $ep$
scattering at HERA and in $\nu p$ $(\overline{\nu} p)$ scattering in
CCFR experiments. We consider all the possible charged weak-current
processes in the perturbative regime. M\'{e}ndez {\it et
al}.\cite{mendez} considered extensively the azimuthal correlation in
leptoproduction. In our paper we analyze the same processes but in
different viewpoints and analyses. Especially we direct our focus on
the experimental aspects since we can now verify the theoretical
results in experiments at HERA or CCFR.

Chay {\it et al}.\cite{chay1} considered the azimuthal asymmetry in
$ep$ scattering with a photon exchange. Here we apply a similar
analysis used in Ref.~\cite{chay1} to charged weak-current processes
in lepton-hadron scattering. The result is striking in the sense that
the final-state particles have a strong azimuthal correlation to the
incoming lepton. We will systematically analyze the azimuthal
asymmetry in this paper. In Sec.~II we briefly review the kinematics
used in lepton-proton scattering. In Sec.~III we define the quantity
$\langle \cos \phi \rangle$ as a measure of the azimuthal correlation
and calculate it to order $\alpha_s$ using perturbative QCD. In
Sec.~IV we analyze numerically the azimuthal correlation in various
processes in which the incoming lepton is an electron, a neutrino, a
positron or an antineutrino. We also compare the results from the
semi-inclusive processes in which we identify a final-state hadron,
say, a charged pion with the results from the inclusive processes in
which we use only the calorimetric information, that is, the energy
and the momentum of each particle (or each jet). In Sec.~V we discuss
the behavior of the azimuthal correlation in each process and the
conclusion is given in Sec.~VI. 

\section{Cross Sections}
Here we briefly review the kinematics in lepton-hadron scattering with
charged weak currents. Let $k_1$ ($k_2$) be the initial (final)
momentum of the incoming (outgoing) lepton, $P_1$ ($P_2$) be the
target (observed final-state hadron) momentum and $p_1$ ($p_2$) be the
incident (scattered) parton momentum. At high energy, the hadrons will
be produced with momenta almost parallel to the virtual $W$-boson
direction, $q^{\mu} = k_1^{\mu} -k_2^{\mu}$. We focus on interactions
that produce nonzero transverse momentum ${\bf P}_{2T}$, perpendicular
to the spatial component of $q^{\mu}$, which we will denote by ${\bf
q}$. We choose the direction of ${\bf q}$ to be the negative $z$
axis. We can write the differential scattering cross section in terms
of the following hadronic variables
\begin{eqnarray}
Q^2 &=& -q^2, \ {\bf P}_T = {\bf P}_{2T}, \ \phi, \nonumber \\
x_H &=& \frac{Q^2}{2P_1\cdot q}, \ y = \frac{P_1 \cdot q}{P_1 \cdot
k_1}, \ z_H = \frac{P_1 \cdot P_2}{P_1 \cdot q},
\label{hadvar}
\end{eqnarray}
and the partonic variables
\begin{equation}
x = \frac{x_H}{\xi} =\frac{Q^2}{2p_1 \cdot q}, \
z=\frac{z_H}{\xi^{\prime}} = \frac{p_1 \cdot p_2}{p_1 \cdot q}.
\label{parvar}
\end{equation}
The azimuthal angle $\phi$ of the outgoing hadron is measured with
respect to ${\bf k}_{1T}$, whose direction is chosen to be the
positive $x$ axis. If we employ jets instead of hadrons, $\phi$ is the 
azimuthal angle of the jet defined by an appropriate jet
algorithm\cite{jetal} and all the hadronic variables are replaced by
the jet variables. 

In the parton model, if we consider the inclusive processes $\ell +
p\rightarrow \ell^{\prime} +X$, in which $\ell$, $\ell^{\prime}$ are
different leptons, the differential cross section is given by
\begin{eqnarray}
\frac{d\sigma}{dx_H dy dz_H d^2 P_T} &=& \sum_i \int dxdz d\xi d^2 p_T 
\delta (x_H -\xi x) \delta(z_H - z) \nonumber \\
&\times&  \delta^{(2)} ({\bf P}_T - {\bf p}_T) F_i (\xi, Q^2)
\frac{d\hat{\sigma}_i}{dx dy dz d^2 p_T} 
\nonumber \\ 
&=& \sum_i \int_{x_H}^1 \frac{dx}{x} \int d^2 p_T \delta^{(2)} ({\bf 
P}_T - {\bf p}_T) F_i (\frac{x_H}{x}, Q^2) \frac{d\hat{\sigma}_i}{dx
dy dz d^2 p_T},
\label{notag} 
\end{eqnarray}
with $d^2 P_T = P_T dP_T d\phi$. The sum $i$ runs over all types of
partons (quarks, antiquarks and gluons) inside the proton and
$d\hat{\sigma}_i$ is the partonic differential cross section. $F_i
(x,Q^2)$ is the parton distribution function of finding the $i$-type
parton inside the proton with the momentum fraction $x$. In
Eq.~(\ref{notag}) we neglect the intrinsic momentum due to the
nonperturbative effects and we identify the momentum of the
final-state hadron (or a jet) with the momentum of the scattered
parton. This approximation is valid if we choose final-state particles
with large transverse momenta. 

If we consider the semi-inclusive process $\ell +
p\rightarrow \ell^{\prime} + h +X$ where $h$ is a detected hadron,
say, a charged pion, the differential cross section is given by
\begin{eqnarray}
\frac{d\sigma}{dx_H dy dz_H d^2P_T} &=& \sum_{ij} \int dx dz
d\xi d\xi^{\prime} d^2p_T \delta (x_H -\xi x) \delta (z_H
-\xi^{\prime} z) \delta^{(2)} ({\bf P}_T - \xi^{\prime} {\bf p}_T)
\nonumber \\ 
&\times& F_i (\xi, Q^2)
\frac{d\hat{\sigma}_{ij}}{dx dy dz d^2 p_T}D_j (\xi^{\prime},Q^2)
\nonumber \\ 
&=& \sum_{ij} \int_{x_H}^1 \frac{dx}{x} \int_{z_H}^1 \frac{dz}{z} \int 
d^2p_T \delta^{(2)} ({\bf P}_T  - \frac{z_H}{z}{\bf p}_T) \nonumber \\ 
&\times&F_i (\frac{x_H}{x}, Q^2) \frac{d\hat{\sigma}_{ij}}{dx dy dz
d^2 p_T}D_j(\frac{z_H}{z},Q^2). 
\label{tagx}
\end{eqnarray}
The sum $i$, $j$ runs over all types of partons. The partonic cross
setion $d\hat{\sigma}_{ij}$ describes the partonic semi-inclusive
process 
\begin{equation}
\ell (k_1) + {\rm parton} \ i (p_1) \rightarrow \ell^{\prime}(k_2) +
{\rm parton} \ j (p_2) +X.
\end{equation}
Here the exchanged gauge boson is a charged $W$ particle. $F_i(x,Q^2)$
is the $i$-type parton distribution function, $D_j(z,Q^2)$ is the
fragmentation function of the $j$-type parton to hadronize into the
observed hadron $h$ with the momentum fraction $z$. These two types of
functions depend on factorization scales and for simplicity we put the
scale to be $Q$, a typical scale in lepton-hadron scattering. 

In order to obtain hadronic cross sections, we have to calculate
partonic cross sections using perturbative QCD. At zeroth order in
$\alpha_s$, the parton cross section for the scattering $\nu +
q\rightarrow e +q^{\prime}$ is given by  
\begin{equation}
\frac{d\hat{\sigma}_q}{dxdydz d^2 p_T} = \frac{G_F^2 m_W^4
|V_{q^{\prime}q}|^2}{\pi} \frac{Q^2}{(Q^2 + m_W^2)^2} \frac{1}{y} 
\delta (1-x) \delta (1-z) \delta^2 ({\bf p}_T),
\label{treeq}
\end{equation}
where $V_{q^{\prime}q}$ is the relevant Cabibbo-Kobayashi-Maskawa
(CKM) matrix element for the process $W^+ +q\rightarrow
q^{\prime}$. $G_F$ is the Fermi constant and $m_W$ is the mass of the
$W$ gauge boson. For the scattering of an antiquark with a neutrino,
$\nu + \overline{q} \rightarrow e +\overline{q}^{\prime}$, the parton
cross section is given by 
\begin{equation}
\frac{d\hat{\sigma}_{\overline{q}}}{dxdydz d^2 p_T} = \frac{G_F^2
m_W^4 |V_{qq^{\prime}}|^2}{\pi} \frac{Q^2}{(Q^2 + m_W^2)^2}
\frac{(1-y)^2}{y}  \delta (1-x) \delta (1-z) \delta^2 ({\bf p}_T).
\label{treeqb}
\end{equation}
The only difference between these cross sections in Eqs.~(\ref{treeq})
and (\ref{treeqb}) is the appearance of the factor $(1-y)^2$. This is
due to the helicity conservation. In short, when particles with the
opposite handedness scatter, we have the factor of $(1-y)^2$ in front,
while it is independent of $y$ when particles with the same handedness
scatter. The cross sections for other processes like $e +q \
(\overline{q}) \rightarrow \nu +q^{\prime}\ (\overline{q^{\prime}})$,
$\overline{\nu} + q\ (\overline{q}) \rightarrow e^+ + q^{\prime} \
(\overline{q}^{\prime})$ and $e^+ + q \ (\overline{q})  \rightarrow
\nu + q^{\prime}\ (\overline{q^{\prime}})$ can be obtained using
crossing symmetries. However since the transverse momentum is zero at
this order, there is no azimuthal correlation at the Born level. 

\unitlength 0.1cm
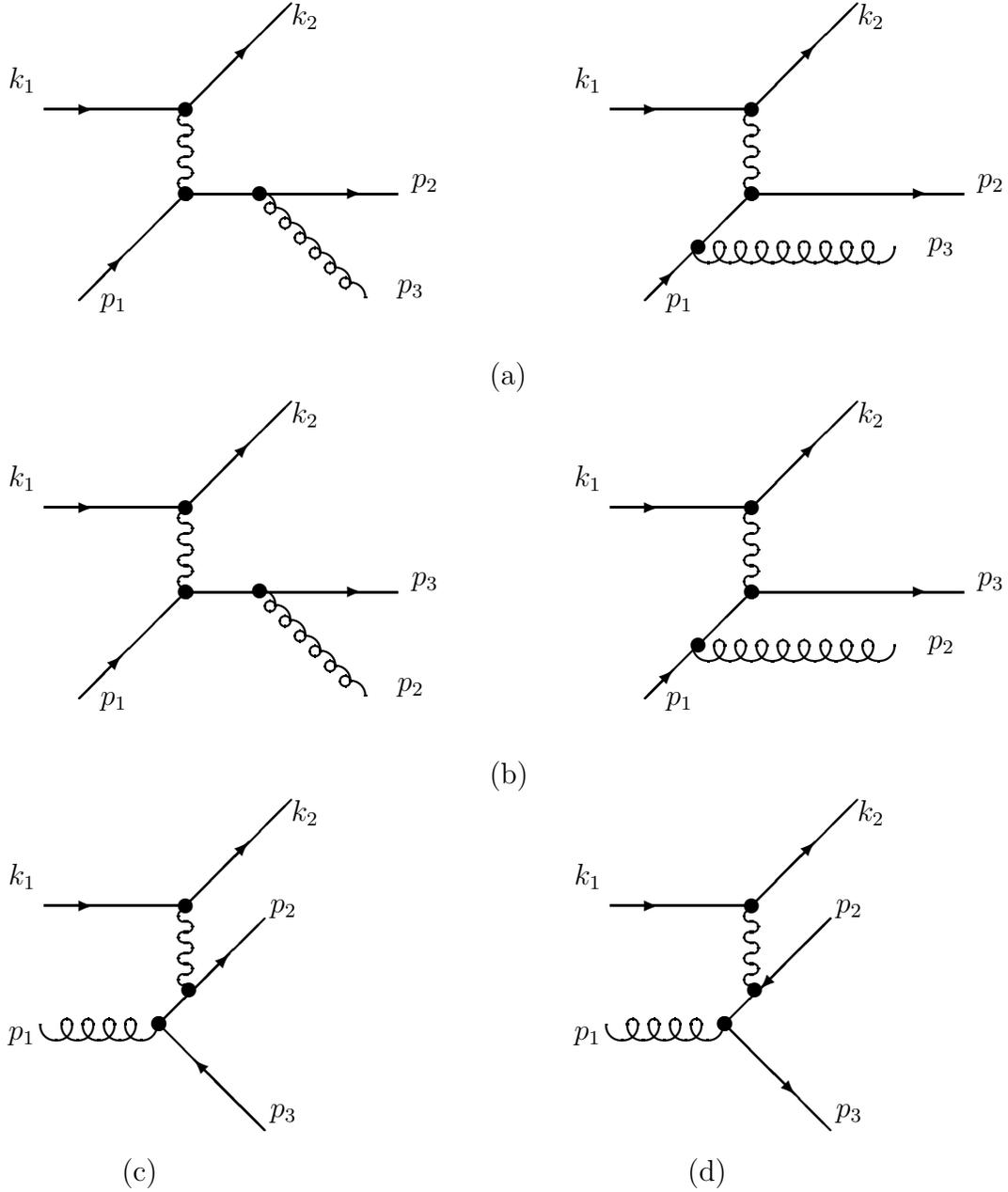
\begin{figure}[b]
\begin{center}
\begin{picture}(160,56)(0,50)
\thicklines
\put(10,93){$k_1$}
\put(15,90){\line(1,0){20}}
\put(15,90){\vector(1,0){7}}
\put(50,102){$k_2$}
\put(35,90){\line(1,1){15}}
\put(35,90){\vector(1,1){9}}
\put(35,90){\circle*{2}}
\multiput(35,88.5)(0,-3){4}{\oval(2,1.5)[l]}
\multiput(35,87)(0,-3){4}{\oval(2,1.5)[r]}
\put(23,62){$p_1$}
\put(20,63){\line(1,1){15}}
\put(20,63){\vector(1,1){6}}
\put(67,79){$p_2$}
\put(35,78){\circle*{2}}
\put(35,78){\line(1,0){30}}
\put(35,78){\vector(1,0){25}}
\put(65,64){$p_3$}
\multiput(45.5,76)(2.121,-2.121){7}{\oval(4.5,4.5)[tr]}
\multiput(47,76)(2.121,-2.121){6}{\oval(1.5,1.5)}
\put(45.5,78.12){\circle*{2}}
\put(90,93){$k_1$}
\put(95,90){\line(1,0){20}}
\put(95,90){\vector(1,0){7}}
\put(115,90){\circle*{2}}
\put(130,102){$k_2$}
\put(115,90){\line(1,1){15}}
\put(115,90){\vector(1,1){9}}
\multiput(115,88.5)(0,-3){4}{\oval(2,1.5)[l]}
\multiput(115,87)(0,-3){4}{\oval(2,1.5)[r]}
\put(103,62){$p_1$}
\put(100,63){\line(1,1){15}}
\put(100,63){\vector(1,1){4}}
\put(147,79){$p_2$}
\put(115,78){\line(1,0){30}}
\put(115,78){\vector(1,0){25}}
\put(115,78){\circle*{2}}
\put(140,70){$p_3$}
\multiput(109,70.5)(3,0){9}{\oval(4.5,4.5)[b]}
\multiput(110.5,70.5)(3,0){8}{\oval(1.5,1.5)[t]}
\put(107.5,70.5){\circle*{2}}
\put(78,51){(a)}
\end{picture}

\begin{picture}(160,56)(0,50)
\thicklines
\put(10,93){$k_1$}
\put(15,90){\line(1,0){20}}
\put(15,90){\vector(1,0){7}}
\put(50,102){$k_2$}
\put(35,90){\line(1,1){15}}
\put(35,90){\vector(1,1){9}}
\put(35,90){\circle*{2}}
\multiput(35,88.5)(0,-3){4}{\oval(2,1.5)[l]}
\multiput(35,87)(0,-3){4}{\oval(2,1.5)[r]}
\put(23,62){$p_1$}
\put(20,63){\line(1,1){15}}
\put(20,63){\vector(1,1){6}}
\put(67,79){$p_3$}
\put(35,78){\circle*{2}}
\put(35,78){\line(1,0){30}}
\put(35,78){\vector(1,0){25}}
\put(65,64){$p_2$}
\multiput(45.5,76)(2.121,-2.121){7}{\oval(4.5,4.5)[tr]}
\multiput(47,76)(2.121,-2.121){6}{\oval(1.5,1.5)}
\put(45.5,78.12){\circle*{2}}

\put(90,93){$k_1$}
\put(95,90){\line(1,0){20}}
\put(95,90){\vector(1,0){7}}
\put(115,90){\circle*{2}}
\put(130,102){$k_2$}
\put(115,90){\line(1,1){15}}
\put(115,90){\vector(1,1){9}}
\multiput(115,88.5)(0,-3){4}{\oval(2,1.5)[l]}
\multiput(115,87)(0,-3){4}{\oval(2,1.5)[r]}
\put(103,62){$p_1$}
\put(100,63){\line(1,1){15}}
\put(100,63){\vector(1,1){4}}
\put(147,79){$p_3$}
\put(115,78){\line(1,0){30}}
\put(115,78){\vector(1,0){25}}
\put(140,70){$p_2$}
\put(115,78){\circle*{2}}
\multiput(109,70.5)(3,0){9}{\oval(4.5,4.5)[b]}
\multiput(110.5,70.5)(3,0){8}{\oval(1.5,1.5)[t]}
\put(107.5,70.5){\circle*{2}}
\put(78,51){(b)}
\end{picture}

\begin{picture}(160,56)(0,50)
\thicklines
\put(10,93){$k_1$}
\put(15,90){\line(1,0){20}}
\put(15,90){\vector(1,0){7}}
\put(50,102){$k_2$}
\put(35,90){\line(1,1){15}}
\put(35,90){\vector(1,1){9}}
\put(35,90){\circle*{2}}
\multiput(35,88.5)(0,-3){4}{\oval(2,1.5)[l]}
\multiput(35,87)(0,-3){4}{\oval(2,1.5)[r]}
\put(35.5,78){\circle*{2}}
\put(47,89){$p_2$}
\put(31.25,73.2){\line(1,1){15}}
\put(31.25,73.2){\vector(1,1){10}}
\put(47,60){$p_3$}
\put(31.25,73.2){\line(1,-1){15}}
\put(46.25,58.2){\vector(-1,1){10}}
\put(31.25,73.2){\circle*{2}}
\put(10,71){$p_1$}
\multiput(16.75,73.2)(3,0){5}{\oval(4.5,4.5)[b]}
\multiput(18.25,73.2)(3,0){4}{\oval(1.5,1.5)[t]}
\put(31.25,73.2){\circle*{2}}
\put(26,51){(c)}

\put(90,93){$k_1$}
\put(95,90){\line(1,0){20}}
\put(95,90){\vector(1,0){7}}
\put(130,102){$k_2$}
\put(115,90){\line(1,1){15}}
\put(115,90){\vector(1,1){9}}
\put(115,90){\circle*{2}}
\multiput(115,88.5)(0,-3){4}{\oval(2,1.5)[l]}
\multiput(115,87)(0,-3){4}{\oval(2,1.5)[r]}
\put(115.5,78){\circle*{2}}
\put(127,89){$p_2$}
\put(111.25,73.2){\line(1,1){15}}
\put(126.25,88.2){\vector(-1,-1){10}}
\put(127,60){$p_3$}
\put(111.25,73.2){\line(1,-1){15}}
\put(111.25,73.2){\vector(1,-1){10}}
\put(111.25,73.2){\circle*{2}}
\put(90,71){$p_1$}
\multiput(96.75,73.2)(3,0){5}{\oval(4.5,4.5)[b]}
\multiput(98.25,73.2)(3,0){4}{\oval(1.5,1.5)[t]}
\put(111.25,73.2){\circle*{2}}
\put(106,51){(d)}
\end{picture}
\end{center}
\caption{Feynman diagrams for charged weak-current processes at
order $\alpha_s$.}
\label{fig1}
\end{figure}

To first order in $\alpha_s$, the parton scattering processes develop
nonzero $p_T$ and nontrivial dependence on the azimuthal angle
$\phi$. The relevant processes are
\begin{eqnarray}
q(p_1)+W^{\pm *} (q) &\rightarrow & q^{\prime}(p_2) + g(p_3),
\label{qq}  \\
q(p_1) +W^{\pm *}(q) &\rightarrow& q^{\prime} (p_3) + g(p_2),
\label{qg} \\ 
\overline{q} (p_1) + W^{\pm *} (q) &\rightarrow &
\overline{q}^{\prime}(p_2) + g(p_3), \label{qbarqbar}  \\
\overline{q}(p_1) +W^{\pm *}(q) &\rightarrow& \overline{q}^{\prime}
(p_3) + g(p_2), 
\label{qbarg} \\ 
g(p_1) + W^{\pm *} (q) &\rightarrow & q(p_2)
+\overline{q}^{\prime}(p_3), \label{gq} \\
g(p_1) + W^{\pm *} (q) &\rightarrow & q(p_3)
+\overline{q}^{\prime}(p_2) \label{gqbar},
\end{eqnarray}
where $g$ is a gluon, $W^{\pm *}$ is the virtual $W$ boson and $q$,
$q^{\prime}$ are quarks. The Feynman diagrams for these processes are
shown in Fig.~\ref{fig1}. Fig.~\ref{fig1}(a) corresponds to
Eq.~(\ref{qq}) [Eq.~(\ref{qbarqbar})] with a quark line (an antiquark 
line) and similarly Fig.~\ref{fig1}(b) corresponds to Eq.~(\ref{qg})
and Eq.~(\ref{qbarg}). Fig.~\ref{fig1}(c) and Fig.~\ref{fig1}(d)
correspond to Eq.~(\ref{gq}) and Eq.~(\ref{gqbar}) respectively. 

Using the Sudakov parametrization we can express $p_2$ in terms of
$x$, $y$ and $z$ as 
\begin{equation}
p_2^{\mu} = \Bigl[(1-x)(1-z)+xz \Bigr] p_1^{\mu} + zq^{\mu}
+\tilde{p}_T^{\mu}, 
\end{equation}
where $\tilde{p}_T = (0, {\bf p}_T, 0)$ is the transverse momentum
with $p_1\cdot \tilde{p}_T = q\cdot \tilde{p}_T =0$. For massless
partons we have
\begin{equation}
p_T^2 = |{\bf p}_{2T}|^2 = \frac{z}{x}(1-x)(1-z)Q^2.
\end{equation}
Similarly we can write
\begin{equation}
k_1^{\mu} =\frac{x}{y}(2-y)p_1^{\mu} +\frac{1}{y} q^{\mu}
+\tilde{k}_T^{\mu},
\end{equation}
with $k_T^2 = (1-y)Q^2/y^2$, where $\tilde{k}_T$ is defined in the
same way as $\tilde{p}_T$. Therefore we have
\begin{equation}
k_1 \cdot p_2 = \frac{Q^2}{2xy} \Bigl[(1-x)(1-z)+xz(1-y)\Bigr] 
-{\bf k}_T \cdot {\bf p}_T,
\end{equation}
and
\begin{equation}
k_2 \cdot p_2 = \frac{Q^2}{2xy} \Bigl[(1-x)(1-y)(1-z) +xz\Bigr] -{\bf
k}_T \cdot {\bf p}_T.
\end{equation}

The semi-inclusive parton scattering cross section for charged
weak-current is given by 
\begin{equation}
\frac{d\hat{\sigma}_{ij}}{dxdydzd^2p_T} = \frac{\alpha_s G_F^2
m_W^4 |V_{q^{\prime}q}|^2}{2\pi^3} \frac{yQ^2}{(Q^2 +m_W^2)^2}  
L_{\mu  \nu}M_{ij}^{\mu\nu} \delta \Bigl(p_T^2
-\frac{z}{x}(1-x)(1-z)Q^2\Bigr),  
\end{equation}
where $L_{\mu\nu}$ is the average squared of the leptonic  charged
current and $M_{ij}^{\mu\nu}$ is the partonic tensor for the incoming
parton $i$ and the outgoing parton $j$. $V_{q^{\prime}q}$ are the CKM
matrix elements. The products $L_{\mu \nu}M_{ij}^{\mu\nu}$ for the
processes in Eqs.~(\ref{qq}), (\ref{qg}), (\ref{qbarqbar}),
(\ref{qbarg}), (\ref{gq}) and (\ref{gqbar}), i.e., $ij=qq$, $qg$, 
$\overline{q}\overline{q}$, $\overline{q}g$, $gq$ and $g\overline{q}$
depend on the types of incoming leptons. For the process $\nu + {\rm
parton} \ i \rightarrow e +{\rm parton} \ j +X$, they are written as 
\begin{eqnarray}
L_{\mu\nu}M_{qq}^{\mu\nu} &=& \frac{4}{3} \frac{(k_1 \cdot p_1)^2 +
(k_2 \cdot p_2)^2}{p_1 \cdot p_3\  p_2 \cdot p_3}, \label{mqq} \\
L_{\mu\nu}M_{qg}^{\mu\nu} &=& \frac{4}{3} \frac{(k_1 \cdot p_1)^2 +
(k_2 \cdot p_3)^2}{p_1 \cdot p_2 \ p_2 \cdot p_3}, \label{mqg}\\
L_{\mu\nu}M_{\overline{q}\overline{q}}^{\mu\nu} &=& \frac{4}{3}
\frac{(k_1 \cdot p_2)^2 + (k_2 \cdot p_1)^2}{p_1 \cdot p_3 \ p_2 \cdot
p_3}, \label{mqbqb}\\  
L_{\mu\nu}M_{\overline{q}g}^{\mu\nu} &=& \frac{4}{3} \frac{(k_1 \cdot
p_3)^2 + (k_2 \cdot p_1)^2}{p_1 \cdot p_2 \ p_2 \cdot p_3},
\label{mqbg}\\ 
L_{\mu\nu}M_{gq}^{\mu\nu} &=& \frac{1}{2}\frac{(k_1 \cdot p_3)^2 +
(k_2 \cdot p_2)^2}{p_1 \cdot p_2 \ p_1 \cdot p_3}, \label{mgq} \\ 
L_{\mu\nu}M_{g\overline{q}}^{\mu\nu} &=& \frac{1}{2}\frac{(k_1 \cdot
p_2)^2 + (k_2 \cdot p_3)^2}{p_1 \cdot p_2\  p_1 \cdot
p_3}. \label{mgqb} 
\end{eqnarray}
Eqs.~(\ref{mqq}) and (\ref{mqg}) correspond to the Feynman diagrams
with quarks in Fig.~\ref{fig1}(a) and \ref{fig1}(b) respectively with
quarks, Eqs.~(\ref{mqbqb}) and (\ref{mqbg}) correspond to the same
diagrams with antiquarks. Eqs.~(\ref{mgq}) and (\ref{mgqb}) correspond
to Fig.~\ref{fig1}(c) and \ref{fig1}(d) respectively. Note that
Eqs.~(\ref{mqg}), (\ref{mqbg}) and (\ref{mgqb}) are obtained from
Eqs.~(\ref{mqq}), (\ref{mqbqb}) and (\ref{mgq}) respectively by
switching $p_2$ and $p_3$. And Eq.~(\ref{mqbqb}) is obtained from
Eq.~(\ref{mqq}) by switching $p_1$ and $p_2$. For the process $e +
{\rm parton}\ i \rightarrow \nu + {\rm parton} \ j+X$, the matrix
elements squared are the same as Eqs.~(\ref{mqq})--(\ref{mgqb}) except 
an extra factor of 1/2 taking into account the spin average of the
incoming electron.   

With the Eqs.~(\ref{mqq})--(\ref{mgqb}), we can also obtain
$L_{\mu\nu} M_{ij}^{\mu\nu}$ for other charged weak-current
processes. For example, for the processes $\overline{\nu} + {\rm
parton} \ i \rightarrow e^+ +{\rm parton} \ j +X$, $L_{\mu\nu}
M_{ij}^{\mu\nu}$ are obtained by switching $k_1$ and $k_2$ in
Eqs.~(\ref{mqq})--(\ref{mgqb}). They are written as  
\begin{eqnarray}
L_{\mu\nu}M_{qq}^{\mu\nu} &=& \frac{4}{3} \frac{(k_2 \cdot p_1)^2 +
(k_1 \cdot p_2)^2}{p_1 \cdot p_3 \ p_2 \cdot p_3}, \label{amqq} \\
L_{\mu\nu}M_{qg}^{\mu\nu} &=& \frac{4}{3} \frac{(k_2 \cdot p_1)^2 +
(k_1 \cdot p_3)^2}{p_1 \cdot p_2 \ p_2 \cdot p_3}, \label{amqg}\\
L_{\mu\nu}M_{\overline{q}\overline{q}}^{\mu\nu} &=& \frac{4}{3}
\frac{(k_2 \cdot p_2)^2 + (k_1 \cdot p_1)^2}{p_1 \cdot p_3 \ p_2 \cdot 
p_3}, \label{amqbqb}\\  
L_{\mu\nu}M_{\overline{q}g}^{\mu\nu} &=& \frac{4}{3} \frac{(k_2 \cdot
p_3)^2 + (k_1 \cdot p_1)^2}{p_1 \cdot p_2 \ p_2 \cdot p_3},
\label{amqbg}\\ 
L_{\mu\nu}M_{gq}^{\mu\nu} &=& \frac{1}{2}\frac{(k_2 \cdot p_3)^2 +
(k_1 \cdot p_2)^2}{p_1 \cdot p_2 \ p_1 \cdot p_3}, \label{amgq} \\ 
L_{\mu\nu}M_{g\overline{q}}^{\mu\nu} &=& \frac{1}{2}\frac{(k_2 \cdot
p_2)^2 + (k_1 \cdot p_3)^2}{p_1 \cdot p_2 \ p_1 \cdot
p_3}. \label{amgqb} 
\end{eqnarray}
By the same argument $L_{\mu\nu}M_{ij}^{\mu\nu}$ for the process $e^+
+{\rm parton} \ i \rightarrow \overline{\nu} +{\rm parton} \ j +X$ are
the same except a factor of 1/2.

\section{Azimuthal Asymmetry}
The azimuthal asymmetry can be characterized by the average value of
$\cos \phi$, which measures the front-back asymmetry of ${\bf P}_{2T}$
along the ${\bf k}_{1T}$ direction. It is defined by
\begin{equation}
\langle \cos \phi \rangle = \frac{\displaystyle \int \Bigl(
d\sigma^{(0)}+d\sigma^{(1)} \Bigr) \cos \phi}{\displaystyle\int \Bigl( 
d\sigma^{(0)} +d\sigma^{(1)}\Bigr)},
\label{cosine}
\end{equation}
where $d\sigma^{(0)}$ ($d\sigma^{(1)}$) is the lowest-order
(first-order in $\alpha_s$) hadronic scattering cross section defined
in Eqs.~(\ref{notag}) or (\ref{tagx}) and the integration over $P_T$,
$\phi$, $x_H$, $y$ and $z_H$ is implied. When we impose a nonzero
transverse momentum cutoff, Eq.~(\ref{cosine}) receives contributions
only from $d\sigma^{(1)}$ both in the numerator and in the
denominator. Note that the zeroth-order cross section is proportional
to $\delta ({\bf P}_T)$. Therefore with the nonzero transverse
momentum cutoff at order $\alpha_s$ in perturbation theory, the
quantity $\langle \cos \phi \rangle$ is independent of $\alpha_s$. 

In fact the azimuthal asymmetry can occur at the Born level if we
include the intrinsic transverse momentum due to the confinement of
partons inside a proton and the fragementation process for partons
into hadrons\cite{cahn,berger,chay1}. However the size of the
intrinsic transverse momentum due to nonperturbative effects is of the
order of a few hundred MeV. Therefore if we make the transverse
momentum cutoff $p_c$ large enough ($\geq 2$ GeV) and choose hadrons
with the transverse momenta larger than $p_c$, we expect that the
contributions from the intrinsic transverse momentum from the
Born-level processes are negligible compared to those from
$\sigma^{(1)}$. In other words the intrinsic transverse momenta of the
partons simply cannot produce hadrons with transverse momenta larger
than $p_c$ and the effects from intrinsic transverse momenta are
suppressed. Therefore, for $p_c$ larger than 2 GeV, 
$\langle \cos \phi\rangle$ is given by, to a good approximation, 
\begin{equation}
\langle \cos \phi \rangle \simeq \frac{\displaystyle \int
d\sigma^{(1)} \cos \phi}{\displaystyle\int d\sigma^{(1)}}.
\label{cosphi}
\end{equation}
In the following analysis we consider $\langle \cos \phi\rangle$ as a
function of the transverse momentum cutoff $p_c$.  

We first consider the azimuthal asymmetry in the inclusive process
$\nu + p \rightarrow e + X$, where $X$ denotes any hadron. The
numerator in Eq.~(\ref{cosphi}) can be written as  
\begin{eqnarray}
\int d\sigma^{(1)} \cos \phi &=& \int d^2 P_T \cos \phi
\frac{d\sigma}{dx_H dy dz_H d^2 P_T} \nonumber \\
&=& \frac{8\alpha_sG_F^2m_W^4}{3\pi^2} \frac{Q^2}{(Q^2 + m_W^2)^2}
\frac{1}{y} \nonumber \\
&\times&\int_{x_H}^1 \frac{dx}{x} (A_{\nu} + B_{\nu} + C_{\nu} +
D_{\nu} + E_{\nu} + F_{\nu}),  
\label{cosnu}
\end{eqnarray}
where
\begin{eqnarray}
A_{\nu}&=& -\sqrt{\frac{(1-y)xz}{(1-x)(1-z)}} \Bigl[ (1-y)(1-x)(1-z)
+xz \Bigr] \nonumber \\
&\times& \Bigl( |V_{ud}|^2 F_d (\frac{x_H}{x}, Q^2) + |V_{cs}|^2 F_s
(\frac{x_H}{x}, Q^2)\Bigr), \nonumber \\
B_{\nu}&=& \sqrt{\frac{(1-y)x(1-z)}{(1-x)z}} \Bigl[ (1-y)(1-x)z
+x(1-z) \Bigr] \nonumber \\
&\times& \Bigl( |V_{ud}|^2 F_d (\frac{x_H}{x}, Q^2) + |V_{cs}|^2 F_s
(\frac{x_H}{x}, Q^2)\Bigr), \nonumber \\
C_{\nu}&=& -\sqrt{\frac{(1-y)xz}{(1-x)(1-z)}} \Bigl[ (1-x)(1-z)
+(1-y)xz \Bigr] \nonumber \\
&\times& \Bigl( |V_{ud}|^2 F_{\overline{u}} (\frac{x_H}{x}, Q^2)
+|V_{cs}|^2 F_{\overline{c}} (\frac{x_H}{x}, Q^2)\Bigr), \nonumber \\
D_{\nu} &=& \sqrt{\frac{(1-y)x(1-z)}{(1-x)z}} \Bigl[ (1-x)z
+ (1-y)x(1-z) \Bigr] \nonumber \\
&\times& \Bigl(|V_{ud}|^2 F_{\overline{u}} (\frac{x_H}{x}, Q^2)
+|V_{cs}|^2 F_{\overline{c}} (\frac{x_H}{x}, Q^2)\Bigr), \nonumber \\
E_{\nu}&=&\frac{3}{8} (1-2x)\sqrt{\frac{(1-y)x(1-x)}{z(1-z)}} \Bigl[ z 
-(1-y)(1-z) \Bigr] \nonumber \\
&\times& \Bigl( |V_{ud}|^2 + |V_{cs}|^2 \Bigr) F_g (\frac{x_H}{x},
Q^2)\Bigr), \nonumber \\ 
F_{\nu}&=&-\frac{3}{8}(1-2x)\sqrt{\frac{(1-y)x(1-x)}{z(1-z)}} \Bigl[
1-z -(1-y)z \Bigr] \nonumber \\
&\times& \Bigl( |V_{ud}|^2 + |V_{cs}|^2 \Bigr) F_g (\frac{x_H}{x},
Q^2)\Bigr).
\label{numnu}
\end{eqnarray}
The denominator can be written as 
\begin{eqnarray}
\int d\sigma^{(1)} &=& \int d^2 P_T \frac{d\sigma}{dx_H dy dz_H d^2
P_T} \nonumber \\
&=& \frac{4\alpha_s G_F^2 m_W^4}{3\pi^2} \frac{Q^2}{(Q^2 + m_W^2)^2}
\frac{1}{y}\nonumber \\
&\times& \int_{x_H}^1 \frac{dx}{x} (A^{\prime}_{\nu} +
B^{\prime}_{\nu} +C^{\prime}_{\nu} +D^{\prime}_{\nu} +E^{\prime}_{\nu}
+F^{\prime}_{\nu}),
\label{cosde}
\end{eqnarray}
where
\begin{eqnarray}
A_{\nu}^{\prime} &=& \Bigl[ \frac{1+x^2 z^2}{(1-x)(1-z)} +4(1-y)xz
+(1-y)^2 (1-x)(1-z) \Bigr] \nonumber \\
&\times& \Bigl( |V_{ud}|^2 F_d (\frac{x_H}{x}, Q^2) + |V_{cs}|^2 F_s
(\frac{x_H}{x}, Q^2)\Bigr), \nonumber \\
B_{\nu}^{\prime}&=& \Bigl[  \frac{1+x^2 (1-z)^2}{(1-x)z} +4(1-y)x(1-z)
+ (1-y)^2 (1-x)z \Bigr] \nonumber \\
&\times& \Bigl( |V_{ud}|^2 F_d (\frac{x_H}{x}, Q^2) + |V_{cs}|^2 F_s
(\frac{x_H}{x}, Q^2)\Bigr), \nonumber \\
C_{\nu}^{\prime} &=& \Bigl[ (1-y)^2 \frac{1+x^2
z^2}{(1-x)(1-z)} + 4(1-y)xz + (1-x)(1-z) \Bigr]  \nonumber \\
&\times& \Bigl( |V_{ud}|^2 F_{\overline{u}} (\frac{x_H}{x}, Q^2)
+|V_{cs}|^2 F_{\overline{c}} (\frac{x_H}{x}, Q^2)\Bigr), \nonumber \\
D_{\nu}^{\prime}&=& \Bigl[ (1-y)^2 \frac{1+x^2
(1-z)^2}{(1-x)z}+ 4(1-y)x(1-z)+ (1-x)z \Bigr] \nonumber \\
&\times& \Bigl( |V_{ud}|^2 F_{\overline{u}} (\frac{x_H}{x}, Q^2)
+|V_{cs}|^2 F_{\overline{c}} (\frac{x_H}{x}, Q^2)\Bigr), \nonumber \\
E_{\nu}^{\prime}&=& \frac{3}{8} \Bigl[ \Bigl( z^2 + (1-y)^2 (1-z)^2 
\Bigr) \frac{x^2 + (1-x)^2}{z(1-z)} + 8(1-y)x(1-x) \nonumber \\
&\times& \Bigl( |V_{ud}|^2 + |V_{cs}|^2 \Bigr) F_g (\frac{x_H}{x},
Q^2), \nonumber \\
F_{\nu}^{\prime}&=& \frac{3}{8} \Bigl[ \Bigl( (1-z)^2 +  (1-y)^2 z^2
\Bigr) \frac{x^2 + (1-x)^2}{z(1-z)} + 8(1-y)x(1-x) \Bigr] \nonumber \\
&\times& \Bigl( |V_{ud}|^2 + |V_{cs}|^2 \Bigr) F_g (\frac{x_H}{x},
Q^2). 
\label{dennu}
\end{eqnarray}
The above six terms in Eqs.~(\ref{numnu}) and (\ref{dennu}) are
obtained from the matrix elements in Eqs.~(\ref{mqq})--(\ref{mgqb})
respectively. For the inclusive process $e + p\rightarrow \nu+X$, the
corresponding quantities are the same except that the quark flavors
are switched, $u\leftrightarrow d$ and $c\leftrightarrow s$ in the
parton distributions functions. There should also be a factor 1/2 from
the incoming electron spin average. However it appears both in the
numerator and in the denominator, hence cancels out.

Now consider the inclusive process $\overline{\nu} + p \rightarrow e^+
+X$. The numerator and the denominator in defining 
$\langle \cos \phi\rangle$ as in Eqs.~(\ref{cosnu}) and (\ref{cosde})
are given by 
\begin{eqnarray}
A_{\overline{\nu}}&=& -\sqrt{\frac{(1-y)xz}{(1-x)(1-z)}} \Bigl[
(1-x)(1-z) +(1-y)xz \Bigr] \nonumber \\
&\times& \Bigl( |V_{ud}|^2 F_u (\frac{x_H}{x}, Q^2) + |V_{cs}|^2 F_c
(\frac{x_H}{x}, Q^2)\Bigr), \nonumber \\
B_{\overline{\nu}}&=& \sqrt{\frac{(1-y)x(1-z)}{(1-x)z}} \Bigl[ (1-x)z
+(1-y)x(1-z) \Bigr] \nonumber \\
&\times& \Bigl( |V_{ud}|^2 F_u (\frac{x_H}{x}, Q^2) + |V_{cs}|^2 F_c
(\frac{x_H}{x}, Q^2)\Bigr), \nonumber \\
C_{\overline{\nu}}&=& -\sqrt{\frac{(1-y)xz}{(1-x)(1-z)}} \Bigl[
(1-y)(1-x)(1-z) +xz \Bigr] \nonumber \\
&\times& \Bigl( |V_{ud}|^2 F_{\overline{d}} (\frac{x_H}{x}, Q^2)
+|V_{cs}|^2 F_{\overline{d}} (\frac{x_H}{x}, Q^2)\Bigr), \nonumber \\
D_{\overline{\nu}} &=& \sqrt{\frac{(1-y)x(1-z)}{(1-x)z}} \Bigl[
(1-y)(1-x)z +x(1-z) \Bigr] \nonumber \\
&\times& \Bigl(|V_{ud}|^2 F_{\overline{d}} (\frac{x_H}{x}, Q^2)
+|V_{cs}|^2 F_{\overline{s}} (\frac{x_H}{x}, Q^2)\Bigr), \nonumber \\
E_{\overline{\nu}}&=&-\frac{3}{8}(1-2x)\sqrt{\frac{x(1-x)(1-y)}{z(1-z)}}
\Bigl[ 1-z -z(1-y) \Bigr] \nonumber \\
&\times& \Bigl( |V_{ud}|^2 + |V_{cs}|^2 \Bigr) F_g (\frac{x_H}{x},
Q^2), \nonumber \\  
F_{\overline{\nu}}&=&\frac{3}{8}
(1-2x)\sqrt{\frac{(1-y)x(1-x)}{z(1-z)}} \Bigl[ z -(1-y)(1-z) \Bigr]
\nonumber \\ 
&\times& \Bigl( |V_{ud}|^2 + |V_{cs}|^2 \Bigr) F_g (\frac{x_H}{x},
Q^2), 
\label{numbnu4}
\end{eqnarray}
and
\begin{eqnarray}
A_{\overline{\nu}}^{\prime} &=& \Bigl[(1-y)^2 \frac{1+x^2
z^2}{(1-x)(1-z)} + 4(1-y)xz + (1-x)(1-z) \Bigr] \nonumber \\
&\times& \Bigl( |V_{ud}|^2 F_u (\frac{x_H}{x}, Q^2) + |V_{cs}|^2 F_c
(\frac{x_H}{x}, Q^2)\Bigr), \nonumber \\
B_{\overline{\nu}}^{\prime}&=& \Bigl[ (1-y)^2 \frac{1+x^2
(1-z)^2}{(1-x)z} + 4(1-y)x(1-z) + (1-x)z  \Bigr] \nonumber \\
&\times& \Bigl( |V_{ud}|^2 F_u (\frac{x_H}{x}, Q^2) + |V_{cs}|^2 F_c
(\frac{x_H}{x}, Q^2)\Bigr), \nonumber \\
C_{\overline{\nu}}^{\prime} &=& \Bigl[ \frac{1+x^2 z^2}{(1-x)(1-z)}
+4(1-y)xz + (1-y)^2 (1-x)(1-z) \Bigr] \nonumber \\
&\times& \Bigl( |V_{ud}|^2 F_{\overline{d}} (\frac{x_H}{x}, Q^2)
+|V_{cs}|^2 F_{\overline{s}} (\frac{x_H}{x}, Q^2)\Bigr), \nonumber \\
D_{\overline{\nu}}^{\prime}&=& \Bigl[ \frac{1+x^2
(1-z)^2}{(1-x)z} +4(1-y)x(1-z) + (1-y)^2 (1-x)z \Bigr] \nonumber \\
&\times& \Bigl( |V_{ud}|^2 F_{\overline{d}} (\frac{x_H}{x}, Q^2)
+|V_{cs}|^2 F_{\overline{s}} (\frac{x_H}{x}, Q^2)\Bigr), \nonumber \\
E_{\overline{\nu}}^{\prime}&=& \frac{3}{8} \Bigl[ \Bigl( (1-z)^2
+(1-y)^2 z^2  \Bigr) \frac{x^2 + (1-x)^2}{z(1-z)} + 8(1-y)x(1-x)
\Bigr] \nonumber \\ 
&\times& \Bigl( |V_{ud}|^2 + |V_{cs}|^2 \Bigr) F_g (\frac{x_H}{x},
Q^2), \nonumber \\
F_{\overline{\nu}}^{\prime}&=& \frac{3}{8} \Bigl[ \Bigl( z^2 + (1-y)^2
(1-z)^2   \Bigr) 
\frac{x^2 + (1-x)^2}{z(1-z)} + 8(1-y) x(1-x) \Bigr] \nonumber \\
&\times& \Bigl( |V_{ud}|^2 + |V_{cs}|^2 \Bigr) F_g (\frac{x_H}{x},
Q^2). 
\label{denbnu4}
\end{eqnarray}
For the process $e^+ + p \rightarrow \overline{\nu} +X$, the
corresponding quantities are the same as in Eqs.~(\ref{numbnu4}) and
(\ref{denbnu4}) except the switch  of the quark
flavors $u\leftrightarrow d$ and $c\leftrightarrow s$ in the parton
distribution functions. 

We can express $\langle \cos \phi \rangle$ using Eq.~(\ref{tagx}) in
the semi-inclusive processes in which  we identify a final-state
charged pion. For the process $\nu + p \rightarrow e + \pi +X$, the
numerator can be written as  
\begin{eqnarray}
\int d\sigma^{(1)} \cos \phi &=& \frac{8\alpha_s G_F^2 m_W^4}{3\pi^2}
\frac{Q^2}{(Q^2 + m_W^2)^2}\frac{1}{y} \nonumber \\
&\times& \int_{x_H}^1 \frac{dx}{x}
\int_{z_H}^1 \frac{dz}{z} (a_{\nu} + b_{\nu}+c_{\nu} + d_{\nu} +
e_{\nu} +f_{\nu}),
\label{num5}
\end{eqnarray}
and the denominator can be written as
\begin{eqnarray}
\int d\sigma^{(1)}  &=& \frac{4\alpha_s G_F^2 m_W^4}{3\pi^2}
\frac{Q^2}{(Q^2 + m_W^2)^2}\frac{1}{y} \nonumber \\
&\times& \int_{x_H}^1 \frac{dx}{x}
\int_{z_H}^1 \frac{dz}{z} (a^{\prime}_{\nu} +
b^{\prime}_{\nu}+c^{\prime}_{\nu} + d^{\prime}_{\nu} +
e^{\prime}_{\nu} +f^{\prime}_{\nu}).
\label{den5}
\end{eqnarray}
The quantities introduced in Eqs.~(\ref{num5}) and (\ref{den5}) are
given as follows:
\begin{eqnarray}
a_{\nu}&=& -\sqrt{\frac{(1-y)xz}{(1-x)(1-z)}} \Bigl[ (1-y)(1-x)(1-z)
+xz \Bigr] \nonumber \\
&\times& \Bigl( |V_{ud}|^2 F_d (\frac{x_H}{x},
Q^2)D_u^{\pi}(\frac{z_H}{z},Q^2)  + |V_{cs}|^2 F_s
(\frac{x_H}{x}, Q^2)D_c^{\pi}(\frac{z_H}{z},Q^2)\Bigr), \nonumber \\ 
b_{\nu}&=& \sqrt{\frac{(1-y)x(1-z)}{(1-x)z}} \Bigl[ (1-y)(1-x)z
+x(1-z) \Bigr] \nonumber \\
&\times& \Bigl( |V_{ud}|^2 F_d (\frac{x_H}{x}, Q^2) + |V_{cs}|^2 F_s
(\frac{x_H}{x}, Q^2)\Bigr)D_g^{\pi}(\frac{z_H}{z},Q^2), \nonumber \\
c_{\nu}&=& -\sqrt{\frac{(1-y)xz}{(1-x)(1-z)}} \Bigl[ (1-x)(1-z)
+(1-y)xz \Bigr] \nonumber \\
&\times& \Bigl( |V_{ud}|^2 F_{\overline{u}} 
(\frac{x_H}{x}, Q^2)  D_{\overline{d}}^{\pi}(\frac{z_H}{z},Q^2)
+|V_{cs}|^2 F_{\overline{c}} (\frac{x_H}{x}, Q^2)
D_{\overline{s}}^{\pi}(\frac{z_H}{z},Q^2) \Bigr), \nonumber \\ 
d_{\nu} &=& \sqrt{\frac{(1-y)x(1-z)}{(1-x)z}} \Bigl[ (1-x)z +
(1-y)x(1-z) \Bigr] \nonumber \\
&\times& \Bigl(|V_{ud}|^2 F_{\overline{u}} (\frac{x_H}{x}, Q^2)
+|V_{cs}|^2 F_{\overline{c}} (\frac{x_H}{x}, Q^2)\Bigr)
D_g^{\pi}(\frac{z_H}{z},Q^2), \nonumber \\
e_{\nu}&=&\frac{3}{8} (1-2x)\sqrt{\frac{(1-y)x(1-x)}{z(1-z)}} \Bigl[ z 
-(1-y)(1-z) \Bigr] \nonumber \\
&\times& \Bigl( |V_{ud}|^2 D_u^{\pi} (\frac{z_H}{z},Q^2)+
|V_{cs}|^2 D_c^{\pi}(\frac{z_H}{z},Q^2) \Bigr) F_g (\frac{x_H}{x}, 
Q^2) , \nonumber \\ 
f_{\nu}&=&-\frac{3}{8}(1-2x)\sqrt{\frac{(1-y)x(1-x)}{z(1-z)}} \Bigl[
1-z - (1-y)z \Bigr] \nonumber \\
&\times& \Bigl( |V_{ud}|^2 D_{\overline{d}}^{\pi} (\frac{z_H}{z},Q^2)
+|V_{cs}|^2 D_{\overline{s}}^{\pi}(\frac{z_H}{z},Q^2) \Bigr) F_g
(\frac{x_H}{x}, Q^2),
\label{numnu5}
\end{eqnarray}
where $D_i^{\pi}(z_H/z,Q^2)$ is the fragmentation function for the
$i$-type parton to fragment into a charged pion. 

The quantities in the denominator are given by
\begin{eqnarray}
a_{\nu}^{\prime} &=& \Bigl[ \frac{1+x^2 z^2}{(1-x)(1-z)} + 4(1-y)xz
+(1-y)^2 (1-x)(1-z)\Bigr] \nonumber \\
&\times& \Bigl( |V_{ud}|^2 F_d (\frac{x_H}{x}, Q^2)
D_u^{\pi}(\frac{z_H}{z},Q^2) + |V_{cs}|^2 F_s 
(\frac{x_H}{x}, Q^2)D_c^{\pi}(\frac{z_H}{z},Q^2)\Bigr), \nonumber \\
b_{\nu}^{\prime}&=& \Bigl[  \frac{1+x^2 (1-z)^2}{(1-x)z} +4(1-y)x(1-z)
+ (1-y)^2 (1-x)z \Bigr] \nonumber \\
&\times& \Bigl( |V_{ud}|^2 F_d (\frac{x_H}{x}, Q^2) + |V_{cs}|^2 F_s
(\frac{x_H}{x}, Q^2)\Bigr) D_g^{\pi}(\frac{z_H}{z},Q^2), \nonumber \\
c_{\nu}^{\prime} &=& \Bigl[ (1-y)^2 \frac{1+x^2
z^2}{(1-x)(1-z)} + 4(1-y)xz + (1-x)(1-z)\Bigr]  \nonumber \\
&\times& \Bigl( |V_{ud}|^2 F_{\overline{u}} (\frac{x_H}{x}, Q^2)
D_{\overline{d}}^{\pi}(\frac{z_H}{z},Q^2) 
+|V_{cs}|^2 F_{\overline{c}} (\frac{x_H}{x}, Q^2)
D_{\overline{s}}^{\pi}(\frac{z_H}{z},Q^2) \Bigr), \nonumber \\
d_{\nu}^{\prime}&=& \Bigl[ (1-y)^2 \frac{1+x^2
(1-z)^2}{(1-x)z}+ 4(1-y)x(1-z) + (1-x)z \Bigr] \nonumber \\
&\times& \Bigl( |V_{ud}|^2 F_{\overline{u}} (\frac{x_H}{x}, Q^2)
+|V_{cs}|^2 F_{\overline{c}} (\frac{x_H}{x}, Q^2)\Bigr)
D_g^{\pi}(\frac{z_H}{z},Q^2), \nonumber \\
e_{\nu}^{\prime}&=& \frac{3}{8} \Bigl[ \Bigl( z^2 + (1-y)^2 (1-z)^2 
\Bigr) \frac{x^2 + (1-x)^2}{z(1-z)} + 8(1-y)x(1-x) \Bigr] \nonumber \\ 
&\times& \Bigl( |V_{ud}|^2 D_u^{\pi} (\frac{z_H}{z},Q^2)
+ |V_{cs}|^2 D_c^{\pi}(\frac{z_H}{z},Q^2)\Bigr) F_g (\frac{x_H}{x},
Q^2), \nonumber \\
f_{\nu}^{\prime}&=& \frac{3}{8} \Bigl[ \Bigl( (1-z)^2 + (1-y)^2 z^2
\Bigr) \frac{x^2 + (1-x)^2}{z(1-z)} + 8(1-y)x(1-x) \Bigr] \nonumber \\
&\times& \Bigl( |V_{ud}|^2 D_{\overline{d}}^{\pi} (\frac{z_H}{z},Q^2)
 + |V_{cs}|^2 D_{\overline{s}}^{\pi}(\frac{z_H}{z},Q^2) \Bigr) F_g
(\frac{x_H}{x}, Q^2). 
\label{dennu5}
\end{eqnarray}

For the process $e+p\rightarrow \nu + \pi +X$, the corresponding
quantities are the same as in Eqs.~(\ref{numnu5}) and (\ref{dennu5})
except that the quark flavor dependence in the parton distribution
functions and the fragmentation functions should be switched in each
$SU(2)$ weak doublet. We can also express the corresponding quantities
in the processes $\overline{\nu} +p \rightarrow e^+ +\pi +X$ and $e^+
+p\rightarrow \overline{\nu}+\pi +X$ accordingly as in inclusive
processes.  

\section{Numerical Analysis}
Let us consider how $\langle \cos \phi\rangle$ behaves numerically
when the QCD effects at next-to-leading order are included. Note that,
if we choose particles with nonzero transverse momentum, $\langle \cos
\phi \rangle$ is independent of $\alpha_s$ to first order in
$\alpha_s$. Furthermore, if we choose the momentum cutoff $p_c$ large
enough, say, larger than 2 GeV, the contribution of the intrinsic
transverse momentum inside a hadron is negligible. In our analysis we
will show the numerical results for the final-state particles with
$p_c \geq 2$ GeV so that we neglect nonperturbative effects.   

\begin{figure}[h]
\vskip -1.0in
\vbox
    {%
    \centerline
    { \epsfbox{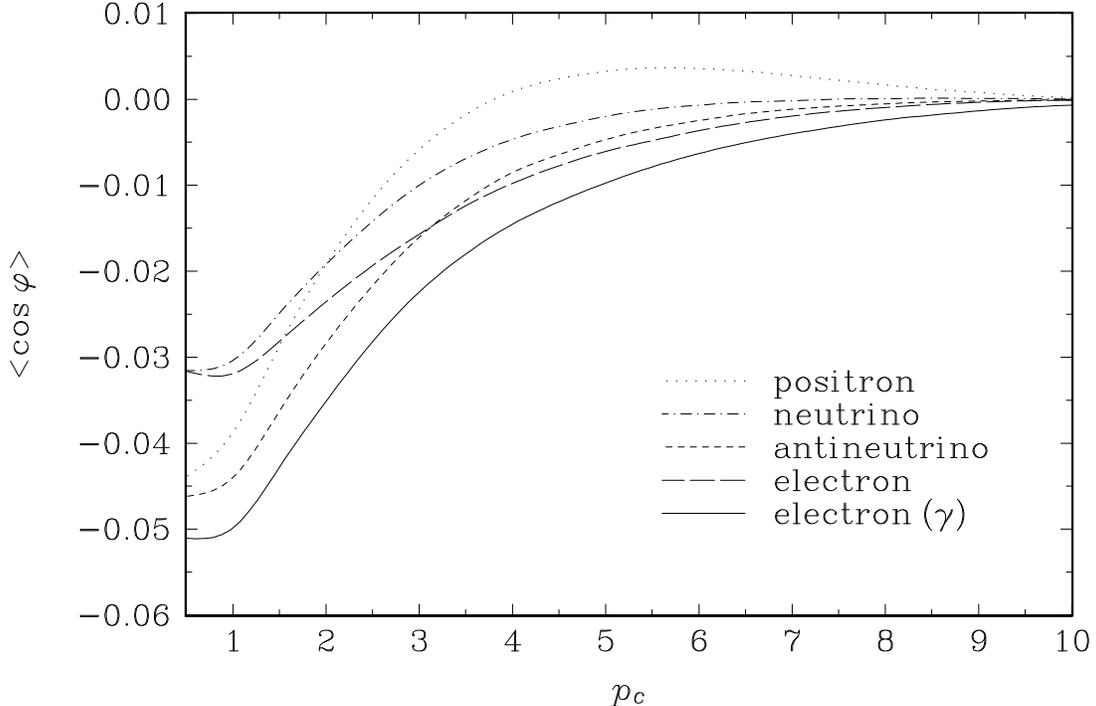} }
\vskip -0.25in
\caption{$\langle \cos\phi\rangle$ versus $p_c$ in inclusive processes.
The leptons listed are the incoming leptons for charged weak-current
processes. The last one with $\gamma$ is from the $ep$ scattering with
a photon exchange.} 
\label{fig2}
    }%
\end{figure}

We show how $\langle \cos \phi \rangle$ behaves as a function of
the transverse momentum cutoff $p_c$ in inclusive processes. The
numerical results for the inclusive processes with different incoming
leptons are listed in Table~1. For comparison we list the result from
the $ep$ scattering in which a photon is exchanged. The plot for
$\langle \cos \phi\rangle$ is shown in Fig.~\ref{fig2}. The numerical
values are obtained by integrating over the ranges $0.05 \leq x_H \leq
0.3$, $0.2\leq y \leq 0.8$ and $0.3 \leq z_H(=z)\leq 1.0$. We also
require that $Q\geq 2$ GeV in order for perturbative QCD to be
valid. We use the Martin-Roberts-Stirling (MRS) (set E) parton
distribution functions\cite{mrse}. 

In Fig.~\ref{fig2} we see that $\langle \cos \phi \rangle$
approaches zero as $p_c$ increases irrespective of the incoming
leptons. If we change kinematic ranges, not only the numerical values
but also the sign change. However the fact that the azimuthal
asymmetry tends to be washed out for large $p_c$ persists. Therefore
the test of perturbative QCD using the azimuthal correlation in
inclusive processes is not feasible until we have better detector 
resolution. However in semi-inclusive processes the situation is
completely different. 

\begin{table}
\caption{$\langle \cos\phi\rangle$ as a function of the transverse
momentum cutoff $p_c$ for inclusive processes. The last column is from
the $ep$ scattering with a photon exchange. The integrated regions
are $0.05 \leq x_H \leq 0.3$, $0.2\leq y \leq 0.8$ and $0.3 \leq
z_H(=z)\leq 1.0$ with $Q\geq 2$ GeV.}    
\begin{center}
\begin{tabular}{cddddd} 
$p_c$ (GeV)&$\nu\rightarrow e$&$e\rightarrow \nu$&
$\overline{\nu}\rightarrow e^+$&  $e^+\rightarrow \overline{\nu}$ &
$e\rightarrow e (\gamma)$ \\ \hline
2.0&$-$0.0192&$-$0.0235&$-$0.0284&$-$0.0192&$-$0.0351 \\ 
3.0&$-$0.0100&$-$0.0157&$-$0.0160&$-$0.00584&$-$0.0224 \\ 
4.0&$-$0.00465&$-$0.00979&$-$0.00852&0.000910&$-$0.0145 \\ 
5.0&$-$0.00200&$-$0.00605&$-$0.00469&0.00325&$-$0.00973 \\ 
6.0&$-$0.000687&$-$0.00364&$-$0.00247&0.00358&$-$0.00632 \\ 
7.0&$-$0.00178&$-$0.00194&$-$0.00116&0.00277&$-$0.00401 \\ 
8.0&6.51$\times 10^{-5}$&$-$0.000952&$-$0.000525&0.00168&$-$0.00239 \\ 
9.0&8.53$\times 10^{-5}$&$-$0.000355&$-$0.000200&0.000784&$-$0.00135
\\  
10.0&2.42$\times 10^{-5}$&$-$0.000107&$-$6.98$\times
10^{-5}$&0.000218&$-$0.000673  
\\ 
\end{tabular}
\end{center}
\end{table}

In the semi-inclusive processes in which we tag a final-state charged
pion, we use analytic fragmentation functions for simplicity. This is
in contrast with studies using Monte Carlo simulation for the
hadronization process\cite{kk}. In our numerical analysis we use
Sehgal's parametrization\cite{sehgal}. Sehgal's parametrization
for the quark fragmentation functions to pions is given by
\begin{equation}
D_j^{\pi} (z) = \frac{1}{z} \Bigl( 0.05 + 1.05 (1-z)^2 \Bigr),
\end{equation}
for $j=u,d,\overline{u},\overline{d}$ and $D_j^{\pi} (z) =0$ for other 
quarks. The gluon fragmentation function to pions is given by
\begin{equation}
D_g^{\pi} (z) = -0.1-2.1z +\frac{2.2}{z} + 4.2 \log z.
\end{equation}
Note that the gluon fragmentation function is ``softer'' than the
quark fragmentation functions, that is, $D_g^{\pi} (z) < D_j^{\pi}
(z)$ for $z>0.21$. This functional form for the gluon is obtained by
assuming that the gluon first breaks up into a quark-antiquark pair,
and then the quarks fragment into the observed hadrons. At large $z$,
the hadrons mainly come from quark fragmentation. For the sake of
simplicity, we also nelgect the QCD-induced scale dependence of these
fragmentation functions. The variation of the fragmentation function
due to the scale dependence largely cancels out in the ratio defining
$\langle \cos \phi\rangle$. 

Since $Q^2 = 2ME_l x_H y$, where $M$ is the proton mass, $E_l$ is the
energy of the incoming lepton in the proton rest frame, when we
integrate over $x_H$ and $y$, the strong coupling constant $\alpha_s
(Q^2)$ should also be included in the integrands in the definition of
$\langle \cos \phi\rangle$. The running coupling constant $\alpha_s$
has the $Q$ dependence as  
\begin{equation}
\alpha_s (Q^2) = \frac{12\pi}{(33-2n_f) \ln (Q^2/\Lambda^2)},
\end{equation}
where $n_f$ is the number of quark flavors whose masses are below
$Q$. However the inclusion of $\alpha_s (Q^2)$ in the integrand is
numerically negligible since it appears both in the numerator and in
the denominator. Therefore in our analysis we do not include
$\alpha_s (Q^2)$ in the integrands. The numerical error in neglecting
the variation of $\alpha_s$ with respect to $Q$ is less than a few
percent.  

\begin{table}
\caption{$\langle \cos\phi\rangle$ as a function of the transverse
momentum cutoff $p_c$ for the semi-inclusive processes with
a final-state charged pion. The last column is from the $ep$
scattering with a photon exchange. The kinematic range is $0.05 \leq 
x_H \leq 0.3$, $0.2\leq y \leq 0.8$ and $0.3 \leq z_H \leq 1.0$ with
$Q\geq 2$ GeV.} 
\begin{center}
\begin{tabular}{cddddd} 
$p_c$ (GeV)&$\nu\rightarrow e$&$e\rightarrow \nu$&
$\overline{\nu}\rightarrow e^+$&  $e^+\rightarrow \overline{\nu}$ &
$e\rightarrow e (\gamma)$ \\ \hline
2.0&$-$0.0515&$-$0.0591&$-$0.115&$-$0.0817&$-$0.0832 \\ 
3.0&$-$0.0443&$-$0.0529&$-$0.128&$-$0.0854&$-$0.0805 \\ 
4.0&$-$0.0399&$-$0.0482&$-$0.146&$-$0.0970&$-$0.0783 \\ 
5.0&$-$0.0364&$-$0.0439&$-$0.166&$-$0.111&$-$0.0762 \\ 
6.0&$-$0.0341&$-$0.0405&$-$0.187&$-$0.127&$-$0.0740 \\ 
7.0&$-$0.0311&$-$0.0366&$-$0.204&$-$0.141&$-$0.0720 \\ 
8.0&$-$0.0289&$-$0.0332&$-$0.219&$-$0.156&$-$0.0698 \\
9.0&$-$0.0256&$-$0.0292&$-$0.224&$-$0.163&$-$0.0660 \\ 
10.0&$-$0.0224&$-$0.0248&$-$0.226&$-$0.173&$-$0.0634 \\
\end{tabular}
\end{center}
\end{table}

\begin{figure}[h]
\vskip -1.0in
\vbox
    {%
    \centerline
    { \epsfbox{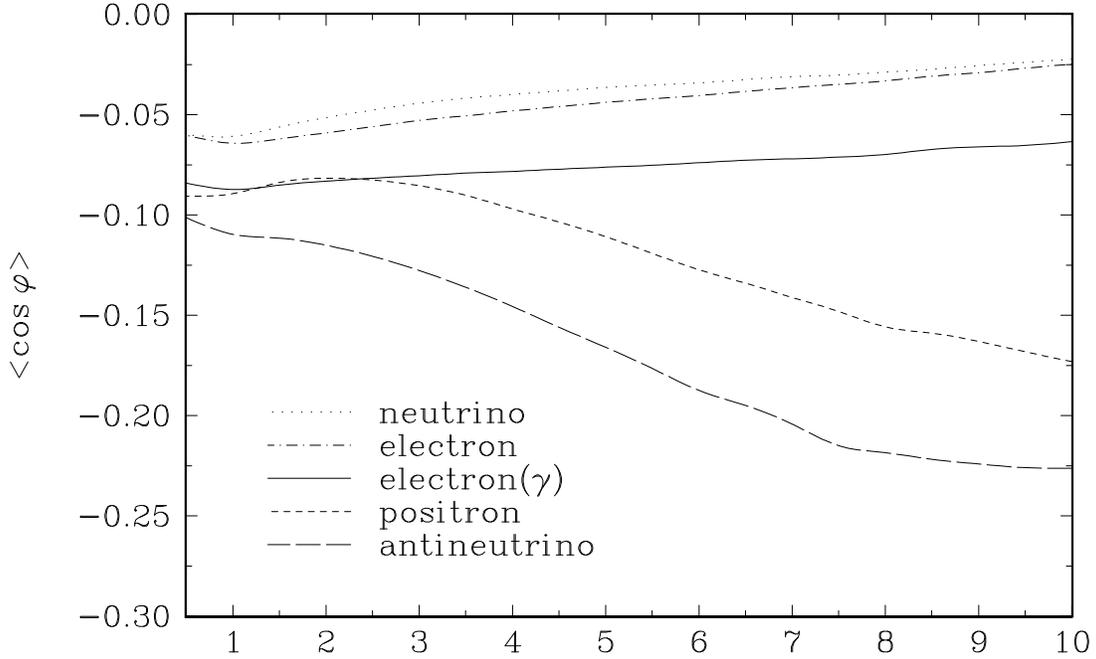} }
\vskip -0.25in
\caption{$\langle \cos\phi\rangle$ versus $p_c$ in semi-inclusive
processes. The leptons listed are the incoming leptons for charged
weak-current processes. The last one with $\gamma$ is from the $ep$
scattering with a photon exchange.}   
\label{fig3}
    }%
\end{figure}

The numerical results for the semi-inclusive processes are given in
Table~2 and the plot is shown in Fig.~\ref{fig3}. The numerical values
are obtained by integrating over the same range as in the analysis of
inclusive processes, $0.05 \leq x_H \leq 0.3$, $0.2\leq y \leq 0.8$
and $0.3 \leq z_H\leq 1.0$ with $Q\geq 2$ GeV. The azimuthal
correlation in semi-inclusive processes shows a rich structure. As
$p_c$ increases, $\langle \cos \phi \rangle$ decreases for the
incoming antineutrino or the positron. On the other hand, for the
incoming neutrino or the electron, it increases and approaches
zero. The result from the $ep$ scattering with a photon exchange is
located between these two cases. This behavior will be analyzed in
detail in the next section and we compare it to the behavior in
inclusive processes.

\section{Discussion}
The most interesting feature of our analysis is the behavior of
$\langle \cos \phi\rangle$ as a function of the transverse momentum
cutoff $p_c$. Let us compare inclusive and semi-inclusive cases shown
in Figs.~\ref{fig2} and \ref{fig3} respectively. In inclusive
processes $\langle \cos \phi \rangle$ approaches zero as $p_c$
increases irrespective of the incoming leptons. On the other hand,
$\langle \cos \phi \rangle$ in semi-inclusive processes is numerically
large compared to that in inclusive processes by an order of magnitude
and it depends on the incoming leptons. However $\langle \cos \phi
\rangle$ remains consistently negative in semi-inclusive
processes. Negative values of $\langle \cos \phi \rangle$ mean that
the final-state particles tend to be emitted to the direction of the
incoming lepton.    

We can understand why there is such asymmetry at order $\alpha_s$ in
the context of color coherence at parton level as noted in
Ref.~\cite{chay1}. When a quark-antiquark pair is produced in a
color-singlet state, soft gluons tend to be emitted inside the cone
defined by the quark-antiquark pair. In our case, we have an incoming
quark and an outgoing quark. However we can regard the incoming quark
as an outgoing antiquark and the pair as a color singlet. Therefore the
configuration in which the outgoing quark is closer to the incoming
lepton and a gluon is emitted between the incoming quark and the
outgoing quark is more probable. It is this configuration that gives
negative $\langle \cos \phi\rangle$ after boosting to the
photon-proton center-of-mass frame assuming that we are in a kinematic
regime where the observed hadron is coming from the fragmentation of
the quark. 

In the semi-inclusive processes in which we identify a final-state
hadron, for example, a charged pion, note that the gluon fragmentation
function is much softer than the quark fragmentation functions. That
is, the gluon fragmentation function $D_g^{\pi} (z_H/z)$ decreases
rapidly as $z_H/z\rightarrow 1$ compared to the quark fragmentation
functions. This is clearly seen in Sehgal's parametrization of the
fragmentation functions. Therefore for large $z_H$ ($z_H \geq 0.3$ in
our numerical result) we effectively pick up the pions which are
fragments of quarks. This is exactly the situation where color
coherence can explain the asymmetry. Of course, final-state quarks can
be produced from the gluon-$W$ fusion. But in this case $\langle \cos
\phi \rangle$ can be either positive or negative, hence there is a
partial cancellation for wide ranges of $x_H$ and $z_H$. 

In inclusive processes, since there appear no fragmentation functions,
both quarks and gluons contribute to the asymmetry. But their
contributions tend to cancel each other since the final-state
particles are emitted in the opposite direction. Note the opposite
signs in the pairs of terms ($A_{\nu}$, $B_{\nu}$), ($C_{\nu}$,
$D_{\nu}$) and ($E_{\nu}$, $F_{\nu}$) in Eq.~(\ref{numnu}). However
the asymmetry can arise depending on the kinematic range. For example,
the valence quarks contribute dominantly for large $x_H/x$ because the
valence quark distribution functions $F_i (x_H/x,Q^2)$ are larger than
other distribution functions. If we compare Figs.~\ref{fig2} and
\ref{fig3}, the cancellation in inclusive processes is illustrated
clearly. The magnitudes of $\langle \cos \phi \rangle$ in inclusive
processes (Fig.~\ref{fig2}) are smaller by an order of magnitude than
those in semi-inclusive processes (Fig.~\ref{fig3}).  
 
Now let us consider the detailed behavior of $\langle \cos \phi
\rangle$ as $p_c$ varies. In evaluating $\langle \cos \phi \rangle$,
there are different combinations of parton distribution functions (and
fragmentation functions in semi-inclusive processes) for different
incoming leptons. However, since these functions appear both in the
denominator and in the numerator, the main difference results from the
matrix elements squared for each process. As the matrix elements
squared for the incoming electron and for the incoming neutrino are
proportional to each other, we expect that the behavior of $\langle
\cos \phi \rangle$ from an incoming electron and from an incoming
neutrino is similar though the magnitudes may be different. This is
true for the cases with an incoming positron and an incoming
antineutrino. This expectation is shown in Fig.~\ref{fig3} for
semi-inclusive processes. It is not clear in Fig.~\ref{fig2} for
inclusive processes since the magnitudes of $\langle \cos \phi\rangle$
are numerically too small to draw any conclusion.

One interesting feature in Fig.~\ref{fig3} is that when the incoming
particle is an antineutrino or a positron, $\langle \cos \phi \rangle$
is more negative compared to the case of the incoming neutrino or
electron. $\langle \cos \phi\rangle$ decreases as $p_c$ increases for
incoming antileptons, while it increases and approaches zero for
incoming leptons. This behavior results from complicated functions
depending on  $x$, $y$, $z$, $x_H$ and $z_H$. Therefore it is
difficult to explain the behavior in a simple way. However we can
explain why $\langle \cos \phi \rangle$ is more negative for incoming
antileptons with large $p_c$. 

In semi-inclusive processes, since we select the hadron with
transverse momentum $P_T$ larger than the transverse momentum cutoff
$p_c$, we have the relation 
\begin{equation}
P_T^2 = \frac{(1-x)(1-z)}{xz}z_H^2 Q^2 = 2 ME_l x_H y
\frac{(1-x)(1-z)}{xz}z_H^2 \geq p_c^2.
\label{cut5}
\end{equation}
The second equality in Eq.~(\ref{cut5}) is obtained by the relation
$Q^2 = 2x_H y ME_l$. For large $p_c$, the phase space is confined to 
the region with small $x$, $z$ and large $x_H$, $y$ and $z_H$. In this 
region the ratio $z_H/z$, which appears in the fragmentation
functions, is large, hence the contribution of the gluon fragmentation
is negligible compared to that of the quark (antiquark)
fragmentation. In other words $b_{\nu}$, $d_{\nu}$ in
Eq.~(\ref{numnu5}) and $b_{\nu}^{\prime}$, $d_{\nu}^{\prime}$ in
Eq.~(\ref{dennu5}) are negligible compared to other
contributions. Similarly large $x_H/x$, which appears in the parton
distribution functions, is preferred hence the contribution of the
distribution functions of sea quarks and gluons is small compared to
that of the valence quark distribution functions since the valence
quark distribution functions are dominant for large $x_H/x$. As a
result $c_{\nu}$, $e_{\nu}$, $f_{\nu}$ terms in (\ref{numnu5}) and
$c_{\nu}^{\prime}$, $e_{\nu}^{\prime}$, $f_{\nu}^{\prime}$ terms in
Eq.~(\ref{dennu5}) are negligible. Therefore $a_{\nu}$ and
$a_{\nu}^{\prime}$ dominate for large $p_c$. It means that the main
contribution to $\langle \cos \phi \rangle$ comes from the scattering
of an initial valence quark into a final-state quark, fragmenting to
the observed pion.

Note that, since the parton distribution functions and the
fragmentation functions appear both in the numerator and in the
denominator, $\langle \cos \phi \rangle$ is mainly affected by
the partonic scattering cross sections, which are functions
of parton variables $x$, $y$ and $z$. For small $x$, $z$ and large
$y$, only the first term in $a_{\nu}^{\prime}$ in the denominator and
the first term in $a_{\nu}$ in the numerator are important. The
partonic part of the integrand in the denominator behaves as
$(xz)^{-1}$ and that in the numerator behaves as
$-(xz)^{-1/2}(1-y)^{3/2}$. Since the integrand in the denominator
grows faster than that of the numerator for small $x$, $z$ and large
$y$, $\langle \cos \phi \rangle$ in semi-inclusive processes
approaches zero for the incoming electron or neutrino for large $p_c$,
but it remains negative.   

In the case of the incoming antineutrino, $a_{\overline{\nu}}$ and
$a_{\overline{\nu}}^{\prime}$ terms are dominant for large $p_c$ as in
the case with the incoming neutrino. But the behavior
of these terms are different. Though we do not present the forms of
$a_{\overline{\nu}}$ and $a_{\overline{\nu}}^{\prime}$ here, we can
see the dependence of $a_{\overline{\nu}}$ and
$a_{\overline{\nu}}^{\prime}$ on the partonic variables $x$, $y$ and
$z$ in Eqs.~(\ref{numbnu4}) and (\ref{denbnu4}) for inclusive
processes since the partonic cross sections are the same. For small
$x$, $z$ and large $y$, only the third term in the denominator 
survives and it behaves as $(xz)^{-1}$. On the other hand, the
integrand in the numerator behaves as
$-(xz)^{-1/2}(1-y)^{1/2}$. Therefore the magnitude of $\langle \cos
\phi\rangle$ is larger than that for the incoming electron or neutrino
by a factor of $(1-y)^{-1}$ in the integrand in the numerator, hence
$\langle \cos \phi \rangle$ is more negative than the case of an
incoming electron or neutrino. In addition, because of this factor
$(1-y)^{-1}$, the difference of $\langle  \cos \phi \rangle$ between
the incoming antineutrino and the incoming positron is larger than
that for the incoming electron and and the incoming neutrino. It is
also interesting to note that the azimuthal asymmetry exhibited by a
photon exchange in the semi-inclusive $ep$ scattering is intermediate
between the two cases in which there are leptons or antileptons. 

The behavior of $\langle \cos \phi \rangle$ in inclusive processes can
be explained by the same argument. In this case we identify the
transverse momentum of the final-state hadron (or a jet) as the
transverse momentum of the scattered parton. It corresponds to setting
$z_H =z$. Therefore we select the final-state particle with the
momentum cutoff $p_c$ satisfying 
\begin{equation}
P_T^2 = \frac{z(1-z)(1-x)}{x} Q^2 = 2ME_l x_H y \frac{z(1-z)(1-x)}{x}
\geq p_c^2.
\end{equation}
Therefore as $p_c$ gets large, the integrated phase space is confined
to a region with small $x$, large $x_H$, $y$ and intermediate $z$ 
between 0 and 1. Since the variable $x_H/x$ in the parton distribution
functions is large, the contribution from the gluon distribution
function is negligible. This means that $E$ and $F$ in
Eqs.~({\ref{numnu}) and (\ref{numbnu4}) and $E^{\prime}$,
$F^{\prime}$ in Eqs.~({\ref{dennu}) and (\ref{denbnu4}) can be
neglected. Therefore remaining $A$, $B$, $C$ and $D$ terms and their
primed quantities contribute to $\langle \cos \phi \rangle$.

As we can see in Eq.~(\ref{cosde}), the integrands in the denominator 
behave as $x^{-1}$ whether the incoming particle is a neutrino or an
antineutrino. In the case of the neutrino, the integrand in the
numerator from $A_{\nu}$, $B_{\nu}$ terms behaves as
$x^{-1/2}(1-y)^{3/2}$, while it behaves as $x^{-1/2}(1-y)^{1/2}$ from
$C_{\nu}$, $D_{\nu}$ terms. These terms are smaller than the
integrands in the denominator. Furthermore there is a partial
cancellation between $A_{\nu}$ and $B_{\nu}$ because they have
opposite signs. This is also true for $C_{\nu}$ and
$D_{\nu}$. Therefore $\langle \cos \phi \rangle$ becomes very
small. The same argument applies to the case of the incoming
antineutrino. 

As $p_c$ gets large, the azimuthal asymmetry tends to be washed
out in inclusive processes. This behavior of $\langle \cos \phi
\rangle$ is expected considering the momentum conservation. In our
case in which there are two outgoing particles in the $W$-proton
frame, the transverse momentum of one particle is balanced by another
particle emitted in the opposite direction. Therefore if we sum over
all the contributions from all the emitted particles, there should be
no azimuthal asymmetry. The small azimuthal asymmetry, as shown in
Fig.~\ref{fig2}, arises since we do not include all the emitted
particles with the given choice of $x_H$, $y$ and $z_H$.  

\section{Conclusion}

We have extensively analyzed the azimuthal correlation of final-state
particles in charged weak-current processes. It is a clean test of
perturbative QCD if we make the transverse momentum cutoff $p_c$
larger than, say, 2 GeV. It turns out that the azimuthal asymmetry is
appreciable in semi-inclusive processes compared to inclusive
processes since the asymmetry mainly comes from the contribution of an
final-state quark due to the soft nature of the gluon fragmentation
function for large $z_H$. In inclusive processes we sum over all the
contributions from quarks (antiquarks) and gluons, and the sum
approaches zero as we include a wider range of variables due to the
momentum conservation.  

In addition the azimuthal asymmetry is more conspicuous for
semi-inclusive processes with an incoming antineutrino or a
positron. Previously there was an attempt to analyze the azimuthal
asymmetry at HERA in $ep$ scattering for electroproduction via a
photon exchange. However since $e^+p$ scattering has been performed at
HERA, we expect that the test of the azimuthal asymmetry is more
feasible because the magnitude of $\langle \cos \phi \rangle$ is
bigger in semi-inclusive processes with an incoming positron. In CCFR
experiments they consider only the inclusive cross section for
$\nu_{\mu}\ (\overline{\nu}_{\mu}) + H \rightarrow \mu \ (\mu^+) +X$, 
where $H$ is the target hadron.  If they are able to identify a
final-state hadron, they will also be able to observe the azimuthal
correlations in various charged weak-current processes.  

The azimuthal asymmetry in lepton-hadron scattering results from a
combination of main ideas in the QCD-improved parton model. As
mentioned above, the parton model states that the hadronic cross
section can be separated into three parts: the parton distribution
functions, the fragmentation functions and the partonic hard
scattering cross section. Each element contributes to the azimuthal
asymmetry. If we make a transverse momentum cutoff $p_c$ large enough
in order for perturbative QCD to be valid, the small-$x$
(large-$x_H/x$) region mainly contributes, hence the contribution from
valence quarks is dominant. At the same time, large $p_c$ implies that
the small-$z$ (large-$z_H/z$) region mainly contributes to the
asymmetry. This means that quark or antiquark fragmentation functions
contribute dominantly. The detailed behavior of $\langle \cos \phi
\rangle$ depends on the hard scattering cross section at parton
level. Therefore the experimental analysis of the azimuthal asymmetry
tests the very basic ideas in the QCD-improved parton model.

\section*{Acknowledgments}
One of the authors (JC) was supported in part by the Ministry of
Education BSRI 96-2408 and the Korea Science and Engineering
Foundation through the SRC program of SNU-CTP and grant No. KOSEF
941-0200-022-2, and the Distinguished Scholar Exchange Program of
Korea Research Foundation. SMK was supported in part by Korea Research 
Foundation.

\end{document}